# Heterobilayers of 2D materials as a platform for excitonic superfluidity

Sunny Gupta[1], Alex Kutana[1], and Boris I. Yakobson[1,2,3]*

[1]Materials Science and NanoEngineering, Rice University, Houston, TX 77005, USA
[2]Department of Chemistry, Rice University, Houston, TX 77005, USA
[3]Smalley-Curl Institute for Nanoscale Science and Technology, Rice University, Houston, TX, 77005 USA

*Corresponding Author, Email: biy@rice.edu

## Abstract

Excitonic condensate has been long-sought within bulk indirect-gap semiconductors, quantum wells, and 2D material layers, all tried as carrying media. Here we propose intrinsically stable 2D semiconductor heterostructures with doubly-indirect overlapping bands as optimal platforms for excitonic condensation. After screening hundreds of 2D materials, we identify candidates where spontaneous excitonic condensation mediated by purely electronic interaction should occur, and hetero-pairs $Sb_2Te_2Se/BiTeCl$, $Hf_2N_2I_2/Zr_2N_2Cl_2$, and $LiAlTe_2/BiTeI$ emerge promising. Unlike monolayers, where excitonic condensation is hampered by Peierls instability, or other bilayers, where doping by applied voltage is required, rendering them essentially non-equilibrium systems, the chemically-specific heterostructures predicted here are lattice-matched, show no detrimental electronic instability, and display broken type-III gap, thus offering optimal carrier density without any gate voltages, in true-equilibrium. Predicted materials can be used to access different parts of electron-hole phase diagram, including BEC-BCS crossover, enabling tantalizing applications in superfluid transport, Josephson-like tunneling, and dissipationless charge counterflow.

## Introduction

Condensation to a macroscopic quantum state is a unique feature of bosonic particles that manifests in macroscopic quantum phenomena as remarkable as superfluidity and superconductivity. Experiments on these systems have affirmed our understanding of quantum theory, stimulated applications, and recent widespread interest in use for quantum computing[1–3]. Apart from the extraordinary transport properties[4,5] of the condensate, the topological excitations in a superfluid are ideal hunting ground for exotic particles[6] such as 't Hooft-Polyakov and Dirac monopoles, or skyrmions, useful for comprehending grand unified theory[7] but also for applications in data storage and spintronics[8,9]. Excitons are composite bosons that are bound states of an electron and hole in a solid and were predicted to undergo condensation[10,11] under appropriate conditions. Compared to commonly known bosonic systems such as atomic gases and liquid $^4$He, excitons have smaller mass and their condensate should remain stable[12] up to higher temperatures[13]. Being created in solids by excitation with light, excitons have a short lifetime—a major obstacle to their condensation.

     A fundamentally different possibility of an excitonic ground state was first discussed by Mott who noted[14] that as valence and conduction bands of a semiconductor start overlapping (e.g., under external pressure) leading to a semi-metallic state (Fig. 1a), Coulomb attraction between the conduction band (CB) electrons and valence band (VB) holes should lead to the spontaneous formation of excitons, delaying the appearance of the metallic state. The semimetal transitions to an excitonic insulator phase, with a $2\Delta$ gap [15,16] in the excitation spectrum, as shown in Fig. 1b.

The latter, also called the Bardeen-Cooper-Schrieffer (BCS) phase, corresponds to the weak coupling limit[11] of composite boson condensate, and holds when $na_0^2 \gg 1$, where $n$ is carrier or pair density and $a_0$ is exciton radius. At low density, $na_0^2 \ll 1$, the screening from carriers is reduced (strong coupling) and the Bose-Einstein condensate (BEC) phase forms. BEC and BCS are different limits[11] of the same boson condensate state.

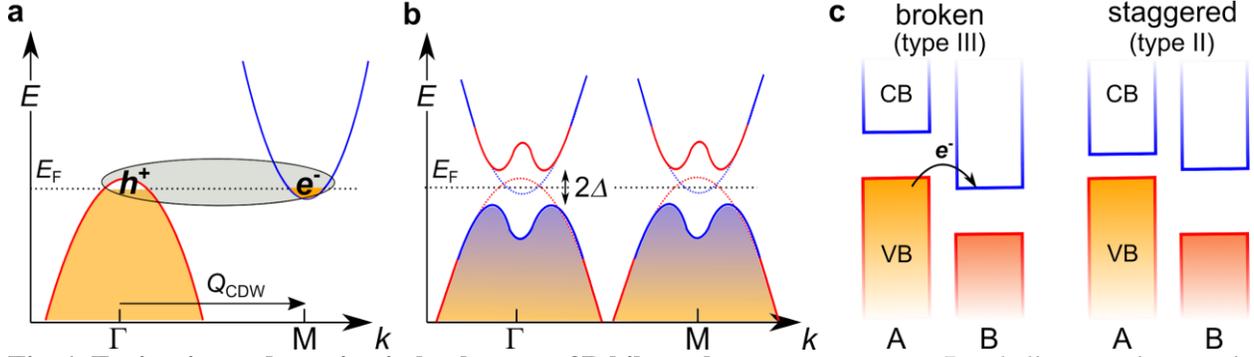

**Fig. 1. Excitonic condensation in broken-gap 2D bilayer heterostructures. a** Band alignment in a semi-metallic system with distinct electron and hole pockets necessary for exciton condensation. **b** For such a system with a negative gap ($E_g$) or when $E_g < E_b$ (the exciton binding energy), the semi-metallic state becomes unstable and opens a gap for an arbitrarily weak electron-hole attraction. The semimetal now transitions to an excitonic insulator, with a gap of $2\Delta$ in the excitation spectrum. **c** For a broken-gap band alignment, the electron is transferred to the conduction band of the acceptor layer B, with a hole remaining in the valence band on the donor layer A. In a staggered-gap alignment, band overlap can be tuned by external perturbation.

After early theoretical works on bulk semi-metallic crystals[10,15,17,18], spatial separation has been proposed[19–21] as a promising way to realize excitonic instability and inverted gap quantum-well bilayer systems were considered. Having electrons and holes spaced apart is deemed essential to reduce many-body interactions that lead to thermodynamic instability near the phase transition[22]. However, in quantum wells typically separated by dozens of Å, electron-hole interaction is weak. Moreover, surface roughness/disorder in quantum wells degrades the coherence of the excitonic state.

Recent interest in excitonic condensation has been renewed by the advent of two-dimensional (2D) materials, providing the advantage of reduced screening and atomically sharp interface. Bilayer graphene has been extensively investigated theoretically[23–26] and even with some experimental evidence[27], however inconclusive, since the phase sought is either not observed or has very low $T_c$; bilayer graphene low effective mass (0.03-0.05 $m_e$)[28,29] results in weak exciton binding and requires intricate device design with electrostatic gating. Moreover, other 2D hetero-bilayers have also been theoretically predicted (ref. [30] and references therein) and recently experimentally demonstrated[31] to exhibit exciton condensation. However, by necessity they require applied voltages for doping with inevitable leakage currents and energy dissipation, in other words all are fundamentally non-equilibrium systems. This is significant in preventing observation of zero-bias Josephson effect[31], not to mention detrimental heating in any device. Recently there was a suggestion of true equilibrium excitonic condensate ground state in 1$T$-TiSe$_2$, a layered 2D semimetal[32,33]. However, TiSe$_2$ is also known to host phonon instability[34,35], possibly making the excitonic phase ground state inaccessible[36] due to lattice reconstruction. In a semimetal, a phonon with wave-vector $Q_{CDW} = \Gamma M$ (Fig. 1a) could render the system unstable by creating a spontaneous crystal distortion (Peierls instability) driven by the electron-phonon interaction, preventing the appearance of the excitonic state[22]. Technically, the mean-field gap equations describing the excitonic insulator and Peierls instability are mathematically identical[37,38,32] and give the same excitation spectrum[16] as shown in Fig. 1b, with the only difference being in the nature of coupling (Coulomb attraction for the former and electron-phonon coupling for the latter), complicating the differentiation of the two phases. Thus, single-layer 2D systems are also potentially plagued by the same problems as bulk crystals - namely, strong interband tunneling transitions and thermodynamic instability, where excitons are mixed in into the ground state of the Peierls insulator[38]. To resolve this issue, both $k$-space and real space separations of the VB and CB are deemed essential for creation of the exciton condensate state.

This work identifies a realistic chemically-specific material system(s), 2D van der Waals (vdW) bilayer heterostructures which are lattice-matched (prerequisite for exciton coherence), do not show detrimental electronic instability (like Peierls), and display broken type-III gap, thus offering the optimal carrier density without any voltage/gating, stable, in true equilibrium. The hetero-bilayers have spatially separated bands and the semimetallic state arises due to difference in the workfunctions of the layers and therefore not requring any electrostatic gating. The electrons and holes are located next to each other on different layers as well as on distinct valleys, their mutual interaction driving excitonic instability and, at lower temperatures, condensation. Since the electron and holes states are spatially separated, the electron-phonon interaction between the layers would be weak, thereby preventing any phonon-driven Peierls instability. Such bilayer heterostructure must be composed of a pair of lattice-matched monolayer 2D semiconductors with either a broken gap (type III) or staggered gap (type II) alignment. For a broken-gap band alignment shown in Fig. 1c, the electron is transferred to the conduction band of the acceptor layer, with a hole remaining in the valence band on the donor layer. Band overlap can also be tuned through external perturbation, as shown in Fig. 1c, starting from the staggered gap lineup.

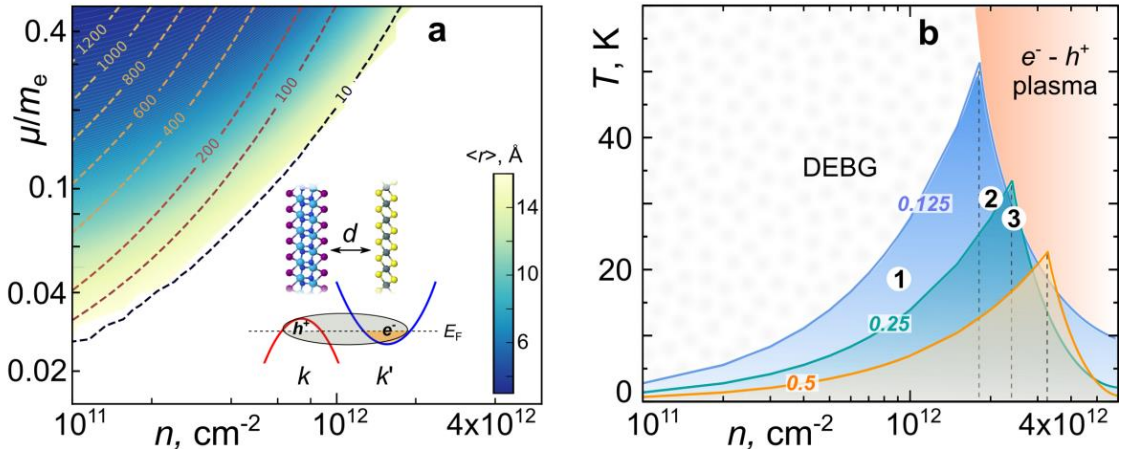

**Fig. 2. Model of exciton condensation in 2D bilayer heterostructure. a** Binding energies (lines labeled by meV) and exciton radius (color map) in a metallic bilayer with interlayer separation $d$=3Å as a function of reduced effective mass $\mu$ and carrier density $n$. Highlighted area shows the region with bound states. **b** Phase diagram for excitons in bilayer semimetal. BKT line for transition to quasi-condensate and BCS line for different reduced mass (0.125, 0.25, and 0.5) are shown (equal carrier masses are assumed). Outside of the condensate region, a degenerate exciton Bose gas (DEBG) exists on the low density side, and electron-hole plasma on the high density side. The estimated critical temperatures ($T_c$) for materials labelled (1, 2, and 3) in Fig. 3 are also shown.

## Results
**Model Hamiltonian analysis of 2D bilayer heterostructure**

In order to guide and accelerate such search of suitable material pairs based on their band alignment, we first perform a model Hamiltonian analysis (see Methods) for estimating the optimal effective mass and carrier density. In a bilayer heterostructure with carrier density $n$, embedded in vacuum and separated by distance $d=3$ Å, the electron and hole states are on different layers and have distinct effective masses. To estimate the binding energies ($E_b$) of individual interlayer excitons in such vdW heterostructures, we adopt the RPA form of the screened potential suitable for bilayers[19,25]. The effective mass Hamiltonian for an electron-hole pair in parabolic bands was constructed and diagonalized using the Gaussian basis set (see Methods for details). Fig. 2a shows the calculated room-temperature binding energies of interlayer excitons as a function of reduced mass $\mu$, and carrier density. Although the potential is always attractive, the model predicts essentially no binding for $n$ higher than ~$10^{12}$ cm$^{-2}$ as well as at carrier masses below $\mu$~0.03 $m_e$.

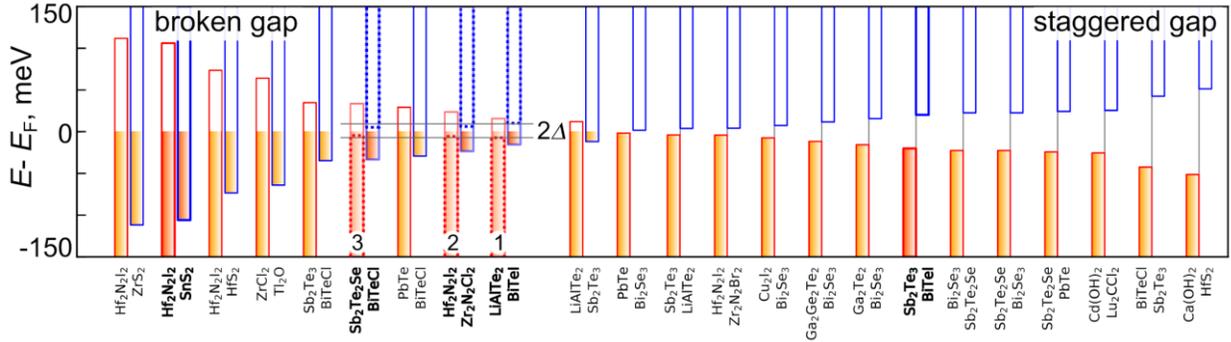

**Fig. 3. Hetero-bilayers band alignments.** Relative positions of valence (left upward columns) and conduction bands (right downward columns) in 2D heterostructures for $\Delta\varepsilon < 2\%$, -0.22 eV $< \Delta E <$ 0.1 eV, $E_g >$ 0.3 eV. The five bilayers considered in detail in this work are highlighted with thicker outline, and the calculated BCS gaps $\Delta$ for heterostructures 1, 2, and 3 are shown.

Favorable conditions for strong interactions and a strongly coupled regime are in the upper left corner of Fig. 2a, i.e. at carrier masses $\mu$~0.5 $m_e$ and densities $n$~$10^{11}$ cm$^{-2}$. At low densities ($na_0^2 \ll 1$) these individual excitons with high binding energy are known[11,12] to condense into a BEC-like phase below a critical temperature. The two-dimensional analog of the three-dimensional BEC phase of excitons is the Berezinskii-Kosterlitz-Thouless (BKT)[39] phase, which, unlike BEC, lacks long range phase coherence, but has local phase coherence and exhibits superfluidity. The temperature $T_{BKT}$ for the BKT superfluid phase transition in a noninteracting 2D Bose liquid is obtained from the universal relation[39,40] $k_B T_{BKT} = \pi n \hbar^2 / 2M$, where $M$ is the exciton mass, and is shown in Fig. 2b for different masses. One should note from Fig. 2a that the binding energies of individual excitons are greater than $k_B T_{BKT}$ at corresponding masses and densities in Fig. 2b, justifying the use of the strongly coupled BKT limit. The dashed lines in Fig. 2b denote the crossover from the BKT phase to the BCS regime where the condensation is best described by the BCS mean field theory[41]. The BCS critical temperature is given by[42] $k_B T_c = 0.57\Delta$ and is shown in Fig. 2b. The order parameter $\Delta$ was evaluated by solving the self-consistent excitonic gap equation for different masses and carrier densities (details in Methods). Overall, Fig. 2b represents a semi-quantitative phase diagram for excitonic particles and denotes optimal $n$ to realize an excitonic condensate phase (see Supplementary Fig. 9 for all other phases). Apart from the phases shown in Fig. 2b, a gas of excitons can also transition to an electron-hole liquid (EHL) state, thereby preventing the realization of an excitonic condensation (BKT/BCS). However, in our bilayer case with electrons and holes spatially separated, the excitons act as oriented electric dipoles with a repulsive interaction between them. The repulsion is important because it prevents the occurrence of EHL state and excitonic BKT/BCS can be realized[43] at low densities. However, at high densities, the EHL state might be more stable[44].

**Identifying optimal 2D bilayer heterostructures**

The densities to realize excitonic condensation should be achievable with existing 2D materials. We search the 2D material database[45] to identify optimal materials by first recalculating and comparing valence and conduction band positions with respect to the vacuum level. The database[45] contains a list of 2D materials which can be exfoliated from experimentally known 3D compounds. Next, after pre-selecting candidate pairs, we calculate the band gaps of 23 heterostructures with broken and staggered band alignments, as shown in Fig. 3. The pairs were chosen so that their lattice mismatch $\Delta\varepsilon <2\%$, and each layer has a band gap >0.3 eV. Band gaps in heterostructures range from -0.22 eV to 0.1 eV, with negative values indicating greater overlap. Assuming effective masses are same, higher band overlap corresponds to higher carrier density, and different heterostructures can be used to access distinct regions of the phase diagram. According to Fig. 2b, $n\sim10^{10}$-$5\times10^{12}$ cm$^{-2}$ is an optimal range of carrier density for excitonic condensation, and the corresponding optimal initial band overlaps are 1 meV<$W_0$<391 meV, assuming $\mu=0.1m_e$, and $d=3$Å (see Methods for details). The final overlap in a heterostructure $W$ is 1 meV<$W$<120 meV. We select three materials pairs $Sb_2Te_2Se/BiTeCl$, $Hf_2N_2I_2/Zr_2N_2Cl_2$, and $LiAlTe_2/BiTeI$ marked in Fig. 3, in which final calculated band overlaps after forming heterostructures are 67 meV, 47 meV, and 38 meV, respectively. Materials with small band gaps can show excitonic instability provided that $E_g<E_b$ and can also in principle exhibit condensation.

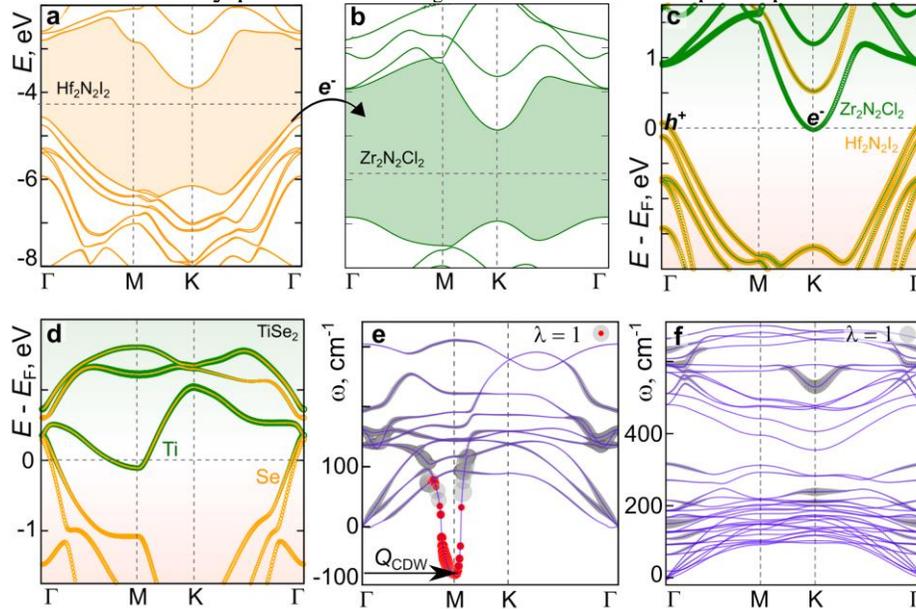

**Fig. 4. Band structures and phonon spectra of 2D $Hf_2N_2I_2$, $Zr_2N_2Cl_2$, $Hf_2N_2I_2/Zr_2N_2Cl_2$, and 1$T$ TiSe$_2$.** Band alignment in **a** $Hf_2N_2I_2$ and **b** $Zr_2N_2Cl_2$. Projected LDA band structures of **c** $Hf_2N_2I_2/Zr_2N_2Cl_2$ heterostructure, and **d** 1$T$ TiSe$_2$. LDA phonon dispersions with electron-phonon couplings $\lambda$ of **e** 1$T$ TiSe$_2$, and **f** $Hf_2N_2I_2/Zr_2N_2Cl_2$ heterostructure. The area of the grey and red circles is proportional to the coupling strength $\lambda$.

We calculate the phonon spectra and band structure of $Hf_2N_2I_2/Zr_2N_2Cl_2$ (see Supplementary Fig. 3 for two other hetero-bilayers), which is instructive to compare with the monolayer 1$T$-TiSe$_2$, well studied[32–35] as a possible candidate for excitonic instability. In Fig. 4a,b the band structures of individual $Hf_2N_2I_2$ and $Zr_2N_2Cl_2$ layers are plotted. Both materials are semiconductors, with the VBM of freestanding $Hf_2N_2I_2$ at -4.57 eV and the CBM of $Zr_2N_2Cl_2$ at -4.88 eV, with respect to the same vacuum level. After a bilayer is formed, the energy overlap between the VBM and CBM decreases but remains, at ~0.05 eV, as seen in Fig. 4c of the $Hf_2N_2I_2/Zr_2N_2Cl_2$ band structure, which also shows that the electron-doped conduction band at the K point is located on the $Zr_2N_2Cl_2$ layer, while hole-doped valence band on the $Hf_2N_2I_2$ layer. As there is little interlayer mixing in either the valence or conduction bands, a bilayer becomes polar. The band structure of $Hf_2N_2I_2/Zr_2N_2Cl_2$ resembles in fact that of monolayer 1$T$-TiSe$_2$ (Fig. 4d), both being indirect-gap semimetals with the VBM and CBM lying at different valleys in the Brillouin zone. Similar to 1$T$-TiSe$_2$, the valence and conduction bands are mainly composed of $p$ and $d$ orbitals, with the VBM originating from iodine $p$ states of $Hf_2N_2I_2$ layer, and smaller contributions from nitrogen $p$ states of $Hf_2N_2I_2$, whereas the CBM originates from zirconium $d$ states, with smaller contributions from $p$ states of N and Cl of $Zr_2N_2Cl_2$ (see Supplementary Fig. 7). Notably, the valence and conduction bands of 1$T$-TiSe$_2$ are strongly coupled to phonons[16,34,46], namely, the bands of TiSe$_2$ are coupled to atomic displacements at modulation vector $Q = \Gamma M$, in Fig. 4e; resulting charge-density wave (CDW) is accompanied by lattice reconstruction. This resembles the Peierls instability which opens the band gap, debatably driving this material to a conventional insulator state[34–36]. In contrast, such instability is absent in $Hf_2N_2I_2/Zr_2N_2Cl_2$, where valence and conduction bands, located at different layers, are decoupled. The phonon spectrum of $Hf_2N_2I_2/Zr_2N_2Cl_2$ in Fig. 4f attests to its stable structure; electron-phonon coupling, largest at K and $\Gamma$ points, is too weak to cause unstable modes. A CDW is expected due to the modulation vector $Q = \Gamma K$[47], however, due to weak electron-phonon coupling, there will not be any lattice distortion. The spontaneous excitonic condensation in this material will be mediated by purely electronic interaction. Thus, excitonic condensation would not be hampered by the phonon instability in a $Hf_2N_2I_2/Zr_2N_2Cl_2$ bilayer or, for that matter, in other vdW bilayers (this was further affirmed by additional supercell calculations, see 3.2 in Supplementary Note 3).

## Discussion

$Hf_2N_2I_2/Zr_2N_2Cl_2$ has a finite modulation vector $Q = \Gamma K$, however, $Sb_2Te_2Se/BiTeCl$, and $LiAlTe_2/BiTeI$ have a zero modulation vector (see Supplementary Fig. 3). This results in band crossing between electron and hole states and also hybridization in $Sb_2Te_2Se/BiTeCl$, and $LiAlTe_2/BiTeI$ (for details see 3.1 in Supplementary Note 3). Single-particle level band hybridization can lead to interband tunneling processes and may fix the phase of the order parameter and destroy superfluidity[47]. However, these tunneling processes can be suppressed by using suitable dielectrics such as $h$BN in between the two layers and superfluidity can be achieved.

From the band structures calculated using LDA functional, we further extract the effective masses (see Supplementary Table 1) and calculate the $n$ in three hetero-bilayers, in order to estimate the critical temperatures for excitonic condensation, marked in Fig. 2b as circled 1, 2 and 3 (see 1.2 in Supplementary Note 1 for the value of $n$ obtained with another functional). We find $n$ to be >$10^{12}$ cm$^{-2}$, with $Hf_2N_2I_2/Zr_2N_2Cl_2$ highest $T_c$ of ~31 K. While the method-dependent spread of $n$ values as estimated by different functionals would yield different estimates for $T_c$, it should not affect the conclusion about the feasibility of excitonic condensation in these heterostructures. Building upon Mott's early ideas[14], external factors such as strain, and possibly gating or changing the interlayer distance (yet not to weaken binding interaction), can fine-tune the carrier density, to access different regions of the phase diagram, and increase the $T_c$ (see Supplementary Note 2 for details). For example, a field of ~1 V/nm can change the carrier density by ~$10^{12}$ cm$^{-2}$, and can quite noticeably shift $T_c$ of materials in the Fig. 2b diagram. Similar perturbation can even be used to tune the band overlap in the hetero-bilayers lying on the flanks of Fig. 3, which were rendered above as suboptimal for excitonic condensation, if those are more readily synthesizable.

In summary, our quantitative analysis demonstrates that $Sb_2Te_2Se$/BiTeCl, $Hf_2N_2I_2$/$Zr_2N_2Cl_2$, $LiAlTe_2$/BiTeI, $Hf_2N_2I_2$/$SnS_2$ and possibly other 2D hetero-bilayers can serve as a platform for realizing an excitonic ground state, in true equilibrium systems (with no need of light-excitation or even static applied voltages). The van der Waals spacing ensures that the lattice thermodynamic stability is preserved, with bands edges being decoupled from the atomic displacements, offering a better material-host for excitonic condensation as compared to recently considered monolayer 1$T$-$TiSe_2$, likely prone to Peierls instability, which hamper exciton condensation. Moreover, the type-III broken gap ensures that no applied gate-voltage is required for doping, contrary to other studied hetero-bilayers. To map the electron-hole system behaviors in a hetero-bilayers, we construct a phase diagram, which for realistic materials suggests excitonic condensation at tens of K at densities $10^{11}$-$10^{13}$ cm$^{-2}$. Various parts of the phase diagram, and most importantly the BEC-BCS crossover can in principle be reached through further tuning by external electric field and strain. We expect the materials predicted here to show superfluid transport properties such as enhanced Josephson-like interlayer tunneling[5], and dissipationless charge counterflow[48], which have been experimentally observed in other bilayer designs and have potential for next-generation electronics and logic devices[49]. Excitons have spin degree of freedom and the spin polarization states of the excitons can also act as a qubit[3]. Moreover, excitonic ferromagnetism[50] may arise from a mix of the singlet and triplet pairing states, which can be used to generate spin currents vital in spintronics. An interesting realization can be possible in nested cylindrical hetero-bilayers, where additional flexoelectric offset[51] must be taken into account.

## Methods

**First principles DFT calculation.** Band structures, phonon frequencies, and electron-phonon coupling coefficients were calculated from density functional theory using LDA functional, as implemented in the Quantum Espresso package[52]. Fully relativistic ultrasoft pseudopotentials were used to represent ionic cores and were obtained from the pslibrary[53] (version 1.0.0, https://dalcorso.github.io/pslibrary/). The pseudopotentials and 60 Ry kinetic energy cutoff was used for all materials except those with Li, Mg, La, Zn, Cu, and Na. For the latter few, we used the cutoffs above the recommended minima, all higher than 60 Ry. A $k$-grid of 81×81×1 was used for Brillouin zone sampling. BFGS algorithm was used for structural relaxation. The structure was relaxed until the Hellmann-Feynman forces on the atoms were less than 2.6 meV/Å. A vacuum of 20 Å was used along direction perpendicular to the 2D sheet to reduce the interaction between the periodic images. Spin-orbit coupling was included in all calculations. The response of the material to an external electric field (Supplementary Fig. 2) was calculated using the Vienna Ab Initio Simulation Package (VASP)[54] implementation. A cutoff of 600 eV for the plane-wave basis was used. Ion-electron interactions were represented by all-electron projector augmented wave potentials. LDA functional is known to fortuitously capture well the interlayer van der Waals interactions[55]. Additionally, semiempirical Grimme's DFT-D2 van der Waals correction to PBE functional was used to estimate the equilibrium interlayer distance and a good agreement with LDA functional was obtained.

We considered all 258 easily-exfoliable layers, with a small unit cell (≤6 atoms/unit cell) from the materials cloud database, as selected by its creators[45]. Among these, 135 were found to have hexagonal lattices, and their pairs-combinations were tested for lattice match. The lattice constant matching is important for excitonic condensation. In a heterostructure with a large difference in lattice constant, the materials are incommensurate, preventing zero-momentum excitons formation and excitonic condensation.

This is analogous to the disappearance of excitonic condensation upon twisting the heterobilayers MoSe$_2$/WSe$_2$, observed recently[31]: upon twisting, the lattices become incommensurate with momentum mismatch between the hole and electron valleys in the materials, the interlayer excitons formed thus have a large momentum, which destroys condensation.

**Model hamiltonian analysis - exciton binding energy.** The model Hamiltonian within the effective mass approximation for an electron-hole pair in a bilayer semimetal is given by

$$H = - \hbar^2/(2\mu_x) \, \partial^2/\partial x^2 - \hbar^2/(2\mu_y) \, \partial^2/\partial y^2 + V_{eh}(r) \qquad (1)$$

Here, $\mu_{x,y}$ are the reduced masses in $x$ and $y$ directions, respectively, and $V_{eh}(r)$ is the effective 2D Coulomb potential in a bilayer metal. Unlike 2D single-layer[56] and bilayer[57] semiconductors, where intralayer and interlayer exciton binding energies are much larger compared to bulk due to weak screening, in semimetallic systems interactions are reduced significantly, leading to weaker binding. Here we adopt the random phase approximation (RPA) form of the screened potential suitable for metallic bilayers[19,25], given by

$$V_{eh}(q) = (2\pi q/\kappa)[e^{-qd}/((q+s)^2 - s^2 e^{-2qd})] \qquad (2)$$

Here, κ is the background dielectric constant assumed to be unity in this work, $s$ is the 2D screening wave number and $d$ is the distance between the bilayers. In the ideal case of a two-band metal, screening $s$ is given by $(gm^*/\kappa)(1-e^{-2\pi\beta n/gm^*})$, where $n$ is the carrier density, $g$ is valley and spin degeneracy, β is $1/k_BT$ and $m^*$ is the effective mass. In a two-dimensional metal with parabolic bands, the screening wave number is independent of density at zero temperature. However, at finite $T$, screening is both density- and temperature-dependent, with increasing temperature leading to reduced screening (for details see 1.1 in Supplementary Note 1); $T$=300 K is assumed here. $V_{eh}(r)$ was obtained from $V_{eh}(q)$ using Fourier transform. The excitonic binding energies $E_b$ and wavefunctions were obtained by finding the eigenvalues and eigenfunctions of the effective mass Hamiltonian (Equation (1)). Gaussian basis set of size 6 was used to expand the ground-state excitonic wavefunction and calculate the matrix elements for the kinetic and potential energy of the Hamiltonian. The expansion was optimized using the downhill simplex algorithm. The parameters for model Hamiltonian analysis were extracted from the heterostructures, not the constituent layers.

**BCS mean field gap equation.** Analogous to the BCS algebra, the order parameter Δ is obtained by solving the below equations self-consistently[21,44]

$$\xi_k = \varepsilon_k - \bar{\mu} - (½)\Sigma_{k'} V_{ee}(|k-k'|)(1-\xi_{k'}/E_{k'}) \qquad (3)$$
$$E_k^2 = \xi_k^2 + \Delta_k^2 \qquad (4)$$
$$\Delta_k = -(½)\Sigma_{k'} V_{eh}(|k-k'|)\Delta_{k'}/E_{k'} \qquad (5)$$

Here, $V_{eh}$ is the attractive Coulomb potential as given by Equation (2), $V_{ee}(q)=(2\pi/\kappa)[(q+s-se^{-2qd})/((q+s)^2-s^2e^{-2qd})]$[30], $\varepsilon_k=½(\hbar^2k^2/2m_e + \hbar^2k^2/2m_h)$, $\bar{\mu}=½(\mu_{ex}-e^2nd/\varepsilon_0\kappa(1+sd))$, $\mu_{ex}$ here is chemical potential for excitons, $n=\Sigma_k ½(1-\xi_k/E_k)$ and $q = |k-k'|$. The order parameter (Δ) in Equation (5) is the gap function. It is zero in the normal state and non-zero in the excitonic insulating state. A value of 2Δ denotes the gap in the excitonic insulator ground state. The form of BCS Equation (5) remains the same when the modulation vector $Q$ is finite[15], as $Q$ does not enter the expression. This is because in $\varepsilon_k$ (which depends on $E_c(k)$ conduction band (electron) dispersion and $E_v(k)$ valence band (hole) dispersion) $k$ is referred to the respective band extrema[15]. We do not include spin degrees of freedom and thus neglect any spin-triplet excitonic order. All quantities in Equations (3,4,5) are assumed to be isotropic i.e. functions of magnitude of $k$, and these equations are solved self-consistently to determine $E_k$ and $\Delta_k$. Subsequently, critical temperature is obtained from the maximum value of $\Delta_k$. The mean field calculations give an approximate depiction of the excitonic condensate phase diagram and quantum Monte Carlo can be appropriate for a much higher accuracy[58].

**Charge transfer model.** For an isolated pair of materials with parabolic bands and VBM-CBM overlap $W_0$, the charge transfer can be estimated using a parallel-plate capacitor model as $n=W_0/e^2(1/C+1/C_q)$ where $C=\varepsilon_0/d$ and $C_q=ge^2\mu/2\pi\hbar^2$ are the classical and quantum bilayer capacitances, respectively. One has $n[10^{13}$ cm$^{-2}]=W_0[eV]/(0.18d[Å]+0.02/\mu[m_e])$ for numerical estimates. In a stacked heterostructure, with a new band overlap $W$, the density is expressed as $n=W(C+C_q)/e^2(1+C/C_q)$, or $n[10^{13}\text{cm}^{-2}]=W[\text{eV}](1+7.6\mu[m_e]d[\text{Å}])/(0.18d[\text{Å}]+0.02/\mu[m_e])$ for numerical estimates.

## Acknowledgement

The authors acknowledge the DoD HPCMP, DOE NERSC, and NSF XSEDE for computing resources.

## Additional information

Supplementary Information accompanies this paper